\documentstyle[preprint,aps,12pt]{revtex}
\begin{document}
\title{
Study of Balance Equations for Hot-Electron
Transport\\ in an Arbitrary Energy Band \ \ (III)}
\author{
Hang-Sheng Wu, \hspace{0.3cm}  Xian-Xiang Huang, \hspace{0.3cm} Ming-Qi Weng}
\address{Department of Physics, University of Science and Technology of
China,\\ Hefei, Anhui, 230026, People's Republic of China\\ \mbox{}}
\maketitle
\draft
\begin{abstract}
By choosing an electron gas resting instead of
drifting in the laboratory \mbox{coordinate} system as the
initial state, the first order perturbation
\mbox{calculation}
 of the previous paper
(Phys. Stat. Sol. (b) {\bf 198}, 785(1996)) is revised and \mbox{extended}
to include the high order field corrections in the expression
for the frictional forces and the energy transfer rates.
The final expressions are \mbox{formally} the same as those
in first order in the electric field,
but the \mbox{distribution}
functions of electrons appearing in them are defined by different expressions. The problems relative to the
distribution function are discussed in detail and a new closed expression
for the distribution function is obtained. The nonlinear impurity-limited resistance of a
strong degenerate electron gas is computed numerically.
The result
calculated by using the new expression for the distribution function is
quite different from that using the displaced Fermi function
when the electric field is sufficiently high.
\end{abstract}
\vskip 2pc
\section{Introduction}
In the previous paper \cite{wu1}, we reformulate the balance equation
theory of Lei and Ting \cite{lt} in {\em ab initio} manner. In this study,
we neither distinguish the degree of freedom of the center of mass from
the relative degree of freedom of the electrons nor adopt the effective
Hamiltonian. All our calculations are carried out in the laboratory
coordinate system and based on the original Hamiltonian ({\em i.e.}
(12) to (21) of \cite{wu1}). We show that the $S$-matrix is, in fact,
also a functional of $H_{E}$, the electrostatic potential of the electrons.
This feature is missing in the Lei-Ting formulation.
It implies that the
expressions of the frictional forces and the energy transfer rate all
depend explicitly on the electric field.
In the second paper \cite{wu2},
this electric field dependence is studied perturbatively. To first order
in the electric field, the results are given by (45), (65), (68) and (69)
of \cite{wu2}.

By choosing a new initial state: both the electron gas and
the phonon gas resting in the laboratory coordinate system,
we revise and extend in the present paper the first order
perturbation calculation given in paper
\cite{wu2} to high order in the electric field. If without a specific
statement, we use in this paper the same notations as those in \cite{wu1}
and \cite{wu2}.

\section{Frictional Forces and Energy Transfer \\ 
Rates: Revision and Extension}
We recall that in the papers \cite{wu1} and \cite{wu2}, we follow
the current hot-electron transport theory \cite{lt}
to assume that at time $t=-\infty$ the system is composed of two
independent equilibrium subsystems: a drifting electron gas with
temperature $T_e$ and a phonon gas with temperature $T$. The corresponding
initial density matrix is
\begin{equation}
\hat{\rho}_0=\hat{\rho}_e\hat{\rho}_{ph}
\label{d-old}
\end{equation}
with
$\hat{\rho}_e={\displaystyle\frac{1}{\Xi_e}}e^{-(\hat{H}_e-\vec{\lambda}\cdot
{\bf p}-\mu {N})/T_e}$
and
$\hat{\rho}_{ph}={\displaystyle\frac{1}{Z_{ph}}}e^{-\hat{H}_{ph}/T}$.
It is pointed out in the Appendix
that in order to keep the consistency of the theory, one should
abandon such initial state from beginning and adopt instead the initial
state: both the electron gas and the phonon gas  resting in the laboratory
coordinate system. And so the initial density matrix (\ref{d-old}) should
be replaced by
\begin{equation}
\hat{\rho}^0=\hat{\rho}_e^0\hat{\rho}_{ph}
\label{d-new}
\end{equation}
with
$\hat{\rho}_e^0={\displaystyle\frac{1}{\Xi_e^0}}e^{-(\hat{H}_e
-\mu {N})/T_e}$.
Keeping this modification in mind, we revise the formulae of \cite{wu2},
and calculate further the high order Feynman diagrams.

In steady state, the balance equations are given by (1) to (5) of
\cite{wu2}. As said above, the density matrix $\hat{\rho}_0$ in these
equations should be replaced by the new one $\hat{\rho}^0$.
Taking (67) of \cite{wu2} into account, we rewrite
these equations as follows
\begin{eqnarray}
&&eN{\bf E}+{\bf f}_{ei}+{\bf f}_{ep}=0 , \label{fb} \\
&&eN{\bf E}\cdot{\bf v}_d+W_{ei}+W_{ep}=0 , \label{eb}
\end{eqnarray}
where
\begin{eqnarray}
{\bf f_{ei}}&=&-{i\over V}\sum_{{\bf k}{\bf q}\sigma}v({\bf q})\,\rho_I({\bf q})
({\bf p_{k+q,\, k+q}-p_{k,\, k}})\, \left\langle S^{+}(t)c_{{\bf k}+{\bf q},\, \sigma}^{\dagger}(t)
c_{\bf k\, \sigma}(t)S(t)\right\rangle,\label{feis}\\
{\bf f_{ep}}&=&-{i\over\sqrt V}\sum_{{\bf kq}\lambda\sigma}M_{{\bf q}\lambda}
({\bf p_{k+q,k+q}-p_{k,k}})\, \left\langle S^{+}(t)\varphi_{{\bf q}\, \lambda}(t)c_{\bf k+q,\, \sigma}^{\dagger}(t)
c_{{\bf k}\sigma}(t)S(t)\right\rangle,\label{feps}\\
W_{ei}&=&-{i\over V}\sum_{{\bf kq}\sigma}v({\bf q})\,\rho_I({\bf q})
(E{({\bf k}+{\bf q})}-E{({\bf k})})\, \left\langle S^{+}(t)c_{{\bf k+q}, \sigma}^{\dagger}(t)
c_{{\bf k}\sigma}(t)S(t)\right\rangle , \label{weis}\\
W_{ep}&=&-{i\over\sqrt V}\sum_{{\bf kq}\lambda\sigma}M_{{\bf q}\lambda}
(E{({\bf k}+{\bf q})}-E{({\bf k})})\, \left\langle S^{+}(t)\varphi_{{\bf q}\, \lambda}(t)c_{\bf k+q,\, \sigma}^{\dagger}(t)
c_{{\bf k}\sigma}(t)S(t)\right\rangle .\label{weps}
\end{eqnarray}
Here $\langle\cdots\rangle$ denotes the statistical average over the density
matrix $\hat{\rho}^0$, {\em i.e.}
\begin{equation}
\langle\cdots\rangle=\mbox{tr}(\cdots\hat{\rho}^0) \ .
\label{av}
\end{equation}
In (\ref{eb}), ${\bf v}_d$ only denotes the drift velocity in the final
steady state.

As shown in \cite{wu2}, the frictional forces (\ref{feis}) and (\ref{feps}),
and the energy transfer rates (\ref{weis}) and (\ref{weps}) can be expressed
in terms of the closed time path Green function
$\underline{G}({\bf k}t,{\bf k}^\prime t^\prime)$. $\underline{G}$ is defined by (17) to
(20) of \cite{wu2} with $\langle\cdots\rangle$ redefined as (\ref{av}).
For example, the frictional force due to impurity scattering, ${\bf f}_{ei}$,
is given by (22) of \cite{wu2}, {\em i.e.}
\begin{equation}
{\bf f}_{ei}=-\frac{1}{V}\sum_{{\bf k}{\bf k}^\prime\sigma}
({\bf p}_{{\bf k}^\prime{\bf k}^\prime}-{\bf p}_{{\bf k}{\bf k}})
\int_{-\infty}^{\infty}\frac{d\omega}{2\pi}\mbox{tr}\Bigl(\hat{L}\underline{G'}({\bf kk}^\prime,\omega)\Bigr)\ .
\label{fei-close}
\end{equation}
The Green function $\underline{G}^\prime$ is calculated by the Keldysh
diagrammatic technique. The Feynman diagrams of order $E^0$ and $E^1$ are
shown in Fig.~1 of \cite{wu2}. The solid line
$\stackrel{{\bf k} \ \omega}
{\longrightarrow\kern -0.6em - \kern -0.6em - \kern -0.6em -}$
represents the
electron propagator
\begin{equation}
\underline{G}^0({\bf k}\omega)=\pmatrix{G_0^R({\bf k}\omega)&G_0^K({\bf k}\omega)\cr
0&G_0^A({\bf k}\omega)\cr}
\label{GM0}
\end{equation}
where
\begin{eqnarray}
G_0^R({\bf k}\omega)&=&\frac{1}{\omega -E{({\bf k})}+
\frac{i}{2\tau{({\bf k})}}}\ ,\\
\label{GR0}
G_0^A({\bf k}\omega)&=&\frac{1}{\omega -E({\bf k})-\frac{i}{2\tau({\bf k})}}\ ,\\
\label{GA0}
G_0^K({\bf k}\omega)&=&\left(G_0^R({\bf k}\omega)-G_0^A({\bf k}\omega)
\right)(1-2f_{\bf k}^0)\ .
\label{GK0}
\end{eqnarray}
They are just (24) to (27) of \cite{wu2}, except that $f_{\bf k}$ in (27)
of \cite{wu2} is replaced by the Fermi function
\begin{equation}
f_{\bf k}^0=\frac{1}{e^{(E{({\bf k})}-\mu)/T_e}+1} \ .
\label{fermi}
\end{equation}
It is easy to see that this revision is the direct consequence of replacing
the initial density matrix $\hat{\rho}_0$ by the new one $\hat{\rho}^0$.
Repeating the calculation expounded in \cite{wu2}, the revised
version of (28), (40) and (41) of \cite{wu2} are easily obtained.
For convenience of latter
discussion, we rewrite them in the form
\begin{eqnarray}
{\bf f}_{ei}^{(0)}&=&
{2\pi \over V}\sum_{{\bf k}{\bf k}^\prime}
n_i|v({\bf k}^\prime-{\bf k})|^2
({\bf p}_{{\bf k}^\prime{\bf k}^\prime}-{\bf p}_{{\bf k}{\bf k}})
L^{(0)}({\bf k},{\bf k}^\prime)
\delta\!\left(E{({\bf k})}-E{({\bf k}^\prime)}\right)\ ,\\ \label{fei0}
{\bf f}_{ei}^{{(1)}}&=&\frac{2\pi}{V}\sum_{{\bf k}{\bf k}^\prime}
n_i|v({\bf k}^\prime-{\bf k})|^{2}({\bf p}_{{\bf k}^\prime{\bf k}^\prime}
-{\bf p}_{{\bf k}{\bf k}})
L^{(1)}({\bf k},{\bf k}^\prime)\delta\!\left(E{({\bf k})} -E{({\bf k}^\prime)}\right)
+\nonumber\\
&+&\frac{2\pi}{V}\sum_{{\bf k}{\bf k}^\prime}
n_i|v({\bf k}^\prime-{\bf k})|^{2}({\bf p}_{{\bf k}^\prime{\bf k}^\prime}
-{\bf p}_{{\bf k}{\bf k}})
L^{(0)}({\bf k},{\bf k}^\prime)C({\bf k},{\bf k}^\prime)
\delta^\prime\!\left(E{({\bf k})} -E{({\bf k}^\prime)}\right)\ ,\label{fei1}
\end{eqnarray}
in which
\begin{eqnarray}
&&L^{(n)}({\bf k},{\bf k}^\prime)=
(-e\tau{({\bf k})}{\bf E}\cdot\nabla_{\bf k})^nf^0_{\bf k}-
(-e\tau{({\bf k}^\prime)}{\bf E}\cdot\nabla_{{\bf k}^\prime})^n
f^0_{{\bf k}^\prime},\\
&&C({\bf k},{\bf k}^\prime)=
-e{\bf E}\cdot\langle{\bf k}|{\bf r}|{\bf k}\rangle
+e{\bf E}\cdot\langle{\bf k}^\prime|{\bf r}|{\bf k}^\prime\rangle .
\end{eqnarray}

The Feynman diagrams of order $E^2$ are given in Fig.~1. Their contributions
to the frictional force ${\bf f}_{ei}$ can be written down by the rules expounded in \cite{wu2}.
We obtain
\begin{eqnarray}
&&\hspace{2.6cm}
{\bf f}_{ei}^{(2)}=-\frac{1}{V}\sum_{{\bf k}{\bf k}^\prime\sigma}
({\bf p}_{{\bf k}^\prime{\bf k}^\prime}
-{\bf p}_{{\bf k}{\bf k}})
n_i|v({\bf k}^\prime-{\bf k})|^{2}
\int\frac{d\omega}{2\pi}\nonumber\\
&\times&
\left\{
\lim_{{{\bf k}_1\rightarrow{\bf k}}
\atop {{\bf k}_1^\prime\rightarrow{\bf k}^\prime}}
\biggl[\left(-e{\bf E}\cdot\langle{\bf k}|{\bf r}|{\bf k}_1\rangle\right)
      \left(-e{\bf E}\cdot\langle{\bf k}_1^\prime|{\bf r}|{\bf k}^\prime\rangle\right)
      \mbox{tr}\!
      \left(\hat{L}\underline{G}^0({\bf k}\omega)
                   \underline{G}^0({\bf k}_1\omega)
                   \underline{G}^0({\bf k}_1^\prime\omega)
                   \underline{G}^0({\bf k}^\prime\omega)
     \right)
\biggr]\right. +\nonumber\\
&+&
\lim_{({\bf k}_1,{\bf k}_2){\bf k}}
\biggl[\left(-e{\bf E}\cdot\langle{\bf k}|{\bf r}|{\bf k}_2\rangle\right)
      \left(-e{\bf E}\cdot\langle{\bf k}_2|{\bf r}|{\bf k}_1\rangle\right)
      \mbox{tr}\!
      \left(\hat{L}\underline{G}^0({\bf k}\omega)
                   \underline{G}^0({\bf k}_2\omega)
                   \underline{G}^0({\bf k}_1\omega)
                   \underline{G}^0({\bf k}^\prime\omega)
     \right)
\biggr]+\nonumber\\
&+&\left.
\lim_{({\bf k}_1^\prime,{\bf k}_2){\bf k}^\prime}
\biggl[\left(-e{\bf E}\cdot\langle{\bf k}_1^\prime|{\bf r}|{\bf k}_2\rangle\right)
      \left(-e{\bf E}\cdot\langle{\bf k}_2|{\bf r}|{\bf k}^\prime\rangle\right)
      \mbox{tr}\!
      \left(\hat{L}\underline{G}^0({\bf k}\omega)
                   \underline{G}^0({\bf k}_1^\prime\omega)
                   \underline{G}^0({\bf k}_2\omega)
                   \underline{G}^0({\bf k}^\prime\omega)
     \right)
\biggr]
\right\}.\nonumber\\
\end{eqnarray}
The symbol $({\bf k}_1,{\bf k}_2){\bf k}_3$ means that ${\bf k}_1$ and
${\bf k}_2$ not only approach each other but also \mbox{approach} to ${\bf k}_3$.
It is reminded that the highly singular matrix element
$\langle{\bf k}_1|{\bf r}|{\bf k}_2\rangle$ should be \mbox{handled} carefully
by the trick expounded in \cite{wu1} and \cite{wu2}. After a lengthy but
straightforward \mbox{calculation}, we finally obtain
\begin{eqnarray}
{\bf f}_{ei}^{(2)}&=&\frac{2\pi}{V}\sum_{{\bf k}{\bf k}^\prime}
({\bf p}_{{\bf k}^\prime{\bf k}^\prime}
-{\bf p}_{{\bf k}{\bf k}})
n_i|v({\bf k}^\prime-{\bf k})|^{2}
L^{(2)}({\bf k},{\bf k}^\prime)\delta\!\left(E{({\bf k})} -E{({\bf k}^\prime)}\right)
+\nonumber\\
&+&\frac{2\pi}{V}
\sum_{{\bf k}{\bf k}^\prime}
({\bf p}_{{\bf k}^\prime {\bf k}^\prime}
-{\bf p}_{{\bf k}{\bf k}})
n_i|v({\bf k}^\prime-{\bf k})|^{2}
L^{(1)}({\bf k},{\bf k}^\prime)C({\bf k},{\bf k}^\prime)
\delta^\prime\!\left(E{({\bf k})} -E{({\bf k}^\prime)}\right)\ .
\label{fei2}
\end{eqnarray}
The same method can be used to calculate any perturbation term. The result is
\begin{eqnarray}
{\bf f}_{ei}^{(n)}&=&\frac{2\pi}{V}\sum_{{\bf k}{\bf k}^\prime}
({\bf p}_{{\bf k}^\prime{\bf k}^\prime}
-{\bf p}_{{\bf k}{\bf k}})
n_i|v({\bf k}^\prime-{\bf k})|^{2}\nonumber\\
&&\times\sum_{j=0}^{n}\frac{1}{j {\lower .5ex\hbox{!}}}L^{(n-j)}({\bf k},{\bf k}^\prime)
[C({\bf k},{\bf k}^\prime)]^{j}
\delta^{(j)}\!\left(E{({\bf k})} -E{({\bf k}^\prime)}\right)\ .
\label{fein}
\end{eqnarray}
$\delta^{(j)}$ denotes the $j$-th derivative of the delta function.
Then the frictional force due to impurity scattering is
$\sum\limits_{n=0}^{\infty}{\bf f}_{ei}^{(n)}$ .
It can be expressed in the following compact and transparent form:
\begin{equation}
{\bf f}_{ei}=\frac{2\pi}{V}\sum_{{\bf k}{\bf k}^\prime}
n_i|v({\bf k}^\prime-{\bf k})|^{2}
({\bf p}_{{\bf k}^\prime {\bf k}^\prime}
-{\bf p}_{{\bf k}{\bf k}})
(\widetilde{f}_{\bf k}-\widetilde{f}_{{\bf k}^\prime})
\delta(\widetilde{E}{({\bf k}^\prime)} -\widetilde{E}{({\bf k})})\ .
\label{fei-last}
\end{equation}
Here,
\begin{eqnarray}
&&\widetilde{E}{({\bf k})}=
E{({\bf k})}-e{\bf E}\cdot\langle{\bf k}|{\bf r}|{\bf k}\rangle\ ,\label{etilde}\\
&&\widetilde{f}_{\bf k}=f_{\bf k}^0+
\sum_{n=1}^\infty(-m{\bf v}{({\bf k})}\cdot\nabla_{\bf k})^n
f_{\bf k}^0\ .
\label{newf}
\end{eqnarray}
where ${\bf v}{({\bf k})}=\frac{1}{m}e{\bf E}\tau{({\bf k})}$.

In the same way, the expression for the frictional force ${\bf f}_{ep}$ and
the energy transfer rates $W_{ei}$ and $W_{ep}$ derived in \cite{wu2} can be
revised and extended to take the contribution of the high order Feynman diagram
into account. The final results are
\begin{eqnarray}
{\bf f}_{ep}&=&\frac{4\pi}{V}\!\!\!\!\!
\sum_{{\bf k}{\bf k}^\prime{\bf q}\lambda\atop
({\bf k}^\prime={\bf k}+{\bf q})}\!\!\!
|M_{{\bf q}\lambda}|^2
({\bf p}_{{\bf k}^\prime{\bf k}^\prime}-{\bf p}_{{\bf k}{\bf k}})\nonumber\\
&&\times[\widetilde{f}_{\bf k}(1-\widetilde{f}_{{\bf k}^\prime})n_{{\bf q}\lambda}
-\widetilde{f}_{{\bf k}^\prime}(1-\widetilde{f}_{\bf k})
(1+n_{{\bf q}\lambda})]
\delta(\widetilde{E}{({\bf k}^\prime)}-\widetilde{E}{({\bf k})}
-\Omega_{{\bf q}\lambda})\ , \label{fep-last} \\
W_{ei}&=&\frac{2\pi}{V}\sum_{{\bf k}{\bf k}^\prime}
n_i|v({\bf k}-{\bf k}^\prime)|^2
(E{({\bf k}^\prime)}-E{({\bf k})})
(\widetilde{f}_{\bf k}-\widetilde{f}_{{\bf k}^\prime})
\delta (\widetilde{E}{({\bf k}^\prime)}-\widetilde{E}{({\bf k})})\ ,
\label{wei-last}\\
W_{ep}&=&\frac{4\pi}{V}\!\!\!
\sum_{{\bf k}{\bf k}^\prime{\bf q}\lambda\atop
({\bf k}^\prime={\bf k}+{\bf q})}\!\!\!\!\!
|M_{{\bf q}\lambda}|^2
(E{({\bf k}^\prime)}-E{({\bf k})})\nonumber\\
&&\times[\widetilde{f}_{\bf k}(1-\widetilde{f}_{{\bf k}^\prime})n_{{\bf q}\lambda}
-\widetilde{f}_{{\bf k}^\prime}(1-\widetilde{f}_{\bf k})
(1+n_{{\bf q}\lambda})]
\delta(\widetilde{E}{({\bf k}^\prime)}-\widetilde{E}{({\bf k})}
-\Omega_{{\bf q}\lambda})\ .
\label{wep-last}
\end{eqnarray}

Comparing the above expressions for ${\bf f}_{ei}$, ${\bf f}_{ep}$,
$W_{ei}$ and $W_{ep}$ with the corresponding ones of \cite{wu2}, we
find that (\ref{fei-last}), (\ref{fep-last}) to (\ref{wep-last}) are
formally the same as those in first order in the electric field
but the distribution functions
appearing in them are defined by different expressions. We further note
that for the special parabolic energy band, the distinction
between $\widetilde{E}({\bf k})$ and $E({\bf k})$ can be neglected
for $\widetilde{E}({\bf k}^\prime)-\widetilde{E}({\bf k})=
E({\bf k}^\prime)-E({\bf k})$. As a result, the energy transfer rate
due to impurity scattering, $W_{ei}$, is equal to zero and (\ref{fei-last}),
(\ref{fep-last}) to (\ref{wep-last}) are identical to those
of Lei and Ting \cite{lt} if the concrete expression
for the distribution function is not concerned.

\section{Distribution Function:
Consistency\\ 
and Parameterization}

In the final steady state, the distribution function of the electron
gas in a uniform electric field
is given by (\ref{newf}).

We first notice the fact that this expression
is an unsatisfactory one and requires \mbox{improving}.
As an example, we
consider the following problem. In weak electric field, we retain only the
first two term of (\ref{newf}), {\em i.e.}
\begin{equation}
\widetilde{f}_{\bf k}=f_{\bf k}^0-e\tau{({\bf k})}{\bf E}\cdot\nabla_{\bf k}
f_{\bf k}^0\ .
\label{1st}
\end{equation}
For simplicity, we consider only impurity scattering and the case where
electrons are governed by parabolic
energy dispersion. The balance equations  of present paper (BEPP), {\em i.e.}
(\ref{fb}), (\ref{eb}), (\ref{fei-last}) and (\ref{fep-last}) to
(\ref{wep-last}), are reduced to $T=T_e$ \cite{lt} and
\begin{equation}
eN{\bf E}=-{\bf f}_{ei}\ .
\label{f1st}
\end{equation}
Using (\ref{1st}), one calculates 
${\bf f}_{ei}$ from (\ref{fei-last}) and obtains
\begin{equation}
-{\bf f}_{ei}=eN{\bf E}\left\langle\frac{\tau{({\bf k})}}{\tau_{tr}{({\bf k})}}
\right\rangle\ ,
\end{equation}
where $\langle A{({\bf k})}\rangle=\frac{4}{3\pi}\frac{1}{V}\sum\limits_{\bf k}
\left(-\frac{\partial f_{\bf k}^0}{\partial E{({\bf k})}}\right)E{({\bf k})}
A{({\bf k})}$ and $\tau_{tr}{({\bf k})}$ is the transport relaxation time
\cite{lt}
\begin{equation}
\frac{1}{\tau_{tr}{({\bf k})}}=
\frac{n_i m}{4\pi k^3}\int_0^{2k}
|v(q)|^2q^3dq\ .
\label{tautr}
\end{equation}
Obviously, both sides of (\ref{f1st}) are not equal. It means that
expression (\ref{newf}) is inconsistent with BEPP. The origin of this
inconsistency is due to the incorrectness of ignoring the renormalization
of the electric field vertex in deriving these equations. Now, let the bare electric
field vertex ({\large\mbox{$\circ$}}) in Fig.~1b and c of \cite{wu2}
be replaced by the renormalized one ($\bullet$) shown in Fig.~2.
We can show that to
first order in electric field,
BEPP is unchanged,
but (\ref{1st}) is modified by replacing $\tau{({\bf k})}$
by a function
$\Lambda_{\bf k}$, {\em i.e.}
\begin{equation}
\widetilde{f}_{\bf k}=f_{\bf k}^0-e\Lambda_{\bf k}{\bf E}\cdot\nabla_{\bf k}
f_{\bf k}^0\ .
\label{1st-rev}
\end{equation}
For the
electron-impurity scattering, $\Lambda_{\bf k}$ is just $\tau_{tr}{({\bf k})}$
\cite{ab,mahan}.
Then, the consistency between the expression of $\widetilde{f}_{\bf k}$ and
BEPP is recovered. Inspired by this success, we replace the bare
electric field vertex in each of the Feynman diagram for
$\underline{G}^\prime$ by the
renormalized one. We have
the same balance equations but a different expression for
$\widetilde{f}_{\bf k}$ which is given by
\begin{equation}
\widetilde{f}_{\bf k}=f_{\bf k}^0+\sum_{n=1}^\infty(-e\Lambda_{\bf k}
{\bf E}\cdot\nabla_{\bf k})^n f_{\bf k}^0\ .
\label{newf-rev}
\end{equation}
$\Lambda_{\bf k}$ is defined by an integral equation \cite{ab,mahan}.
We do not follow this approach here due to
the complexity of the mathematics involved.

We instead follow the approach close to the current hot-electron
transport theory \cite{lt,cornwell} to replace
${\bf v}{({\bf k})}$ by a \mbox{${\bf k}$-independent} one,
${\bf v}$, and therefore write (\ref{newf}) in the form
\begin{equation}
\widetilde{f}_{\bf k}=f_{\bf k}^0+
\sum_{n=1}^\infty(-m{\bf v}\cdot\nabla_{\bf k})^n
f_{\bf k}^0\ .
\label{last}
\end{equation}
The point of this alternate approach is to consider ${\bf v}$
as an unknown parameter and determine it in the following fashion:
combining (\ref{last}) with BEPP and the equations
\begin{eqnarray}
&&2\sum_{\bf k}\widetilde{f}_{\bf k}=N\label{constr1}\ ,\\
&&2\sum_{\bf k}{\bf p}_{\bf kk}\widetilde{f}_{\bf k}=Nm{\bf v}_d\ ,
\label{constr2}
\end{eqnarray}
and solving them self-consistently to get ${\bf v}$, together
with $T_e$, $\mu$ and ${\bf v}_d$ for given $N$, $T$ and ${\bf E}$.
The consistency between the expression for $\widetilde{f}_{\bf k}$
and BEPP is obviously automatically \mbox{satisfied}.
It implies that if ${\bf v}$ is determined in this way,
(\ref{last}) is identical with (\ref{newf-rev}) within the \mbox{approximation}
to neglect the ${\bf k}$-dependent of $\Lambda_{\bf k}$.
So, this alternative approach not only
keep the merit of simplicity in mathematics but also take 
automatically the renormalization of the electric field vertex.

We rewrite the series solution (\ref{last}) compactly in the form
\begin{equation}
\widetilde{f}_{\bf k}=\frac{1}{1+m{\bf v}\cdot\nabla_{\bf k}}
f_{\bf k}^0\ .
\label{cf}
\end{equation}
Multiplying both side by the operator
$1+m{\bf v}\cdot\nabla_{\bf k}$, we get
\begin{equation}
(1+m{\bf v}\cdot\nabla_{\bf k})\widetilde{f}_{\bf k}=f_{\bf k}^0\ .
\label{cf1}
\end{equation}
Then we obtain the following analytical expression for $\widetilde{f}_{\bf k}$
through solving the differential equation (\ref{cf1}):
\begin{equation}
\widetilde{f}_{\bf k}=\int_0^\infty d\xi e^{-\xi} f_{\bf k}(\xi) \ .
\label{int_sol}
\end{equation}
It shows that $\widetilde{f}_{\bf k}$ is equal to a weight average of
$f_{\bf k}(\xi)$ which is defined as
\begin{equation}
f_{\bf k}(\xi)=\frac{1}{\exp [(E_{\bf k}(\xi)-\mu)/T_e]+1}\ ,
\label{cf2}
\end{equation}
with
\begin{equation}
E_{\bf k}(\xi)=E({\bf k}-m{\bf v}\xi)\ .
\label{ekxi}
\end{equation}
The parameter ${\bf v}$ connects with ${\bf v}_d$, the drift velocity of
the electron gas in the final steady state, by (\ref{constr2}).
One can prove from (\ref{constr2}) that
${\bf v}$ is just ${\bf v}_d$ only for the special
parabolic energy band.

\section{Concluding Remarks}
It is remarkable that (\ref{cf1}) is nothing but
a simplified form of the
Boltzmann equation. In fact,
in the relaxation time approximation, the Boltzmann equation of an electron
gas in the presence
of a uniform electric field is
\begin{equation}
\frac{\partial\widetilde{f}_{\bf k}}{\partial t}+
e{\bf E}\cdot\nabla_{\bf k}\widetilde{f}_{\bf k}
=-\frac{\widetilde{f}_{\bf k}-f_{\bf k}^0}{{\tau}_{tr}}\ .
\label{bol}
\end{equation}
${\tau}_{tr}$ is the approximate ${\bf k}$-independent relaxation
time.
In steady state, $\frac{\partial}{\partial t} \widetilde{f}_{\bf k}=0$.
We rewrite (\ref{bol}) in the form
\begin{equation}
\widetilde{f}_{\bf k}+e{\tau}_{tr}{\bf E}\cdot\nabla_{\bf k}\widetilde{f}_{\bf k}
=f_{\bf k}^0\ .
\label{bol2}
\end{equation}
It is just (\ref{cf1}) with ${\bf v}={e{\bf E}{\tau}_{tr}}/{m}$.
Thus (\ref{int_sol}) is a homogeneous steady solution
of the \mbox{Boltzmann} equation in constant relaxation time approximation.
We recall that in the \mbox{approach} of balance equation,
the quantity
${\bf v}$ in (\ref{int_sol}) is considered as an unknown parameter
and determined by solving the complete set of
the equations: (\ref{int_sol}) with BEPP, (\ref{constr1})
and (\ref{constr2}). This trick
of \mbox{parameterization} not only provides a method to choose a proper value
of the relaxation time for given \mbox{temperature} $T$ and electric field ${\bf E}$,
but also ensure
(\ref{int_sol}) consistent with BEPP.

We expand $f_{\bf k}(\xi)$ in the series
\begin{equation}
f_{\bf k}(\xi)=\sum_{n=0}^\infty\frac{1}{n!}
\left.\frac{\partial^n f_{\bf k}(\xi)}{\partial\xi^n}\right|_{\xi=1}(\xi-1)^n\ .
\label{f-exp}
\end{equation}
Substituting it into (\ref{int_sol}) and carrying out the $\xi$-integration,
we have
\begin{equation}
\widetilde{f}_{\bf k}=f_{\bf k}{(1)}+\frac{1}{2}
\left.\frac{\partial^2 f_{\bf k}(\xi)}{\partial\xi^2}\right|_{\xi=1}
+\frac{1}{3}
\left.\frac{\partial^3 f_{\bf k}(\xi)}{\partial\xi^3}\right|_{\xi=1}
+\cdots \ ,
\label{wf-new}
\end{equation}
in which $f_{\bf k}{(1)}$ is just the displaced Fermi function used
by the previous authors \cite{lt,cornwell}, {\em i.e.}
\begin{equation}
f_{\bf k}^{d}=\frac{1}{e^{(E({\bf k}-m{\bf v})-\mu)/T_e}+1} \ .
\label{flt}
\end{equation}
In view of (\ref{cf2}) and
(\ref{ekxi}), we rewrite (\ref{wf-new}) in the form
\begin{equation}
\widetilde{f}_{\bf k}-f_{\bf k}^{d}=\frac{1}{2}
\left[
\left(m{\bf v}\cdot\nabla_{\bf k} E_{\bf k}{(1)}\right)^2
\frac{
\partial^2 f_{\bf k}{(1)}}{\partial E_{\bf k}{(1)}^2}+
\biggl(
\left(m{\bf v}\cdot\nabla_{\bf k}\right)^2\!\! E_{\bf k}{(1)}
\biggr)
\frac{\partial f_{\bf k}{(1)}}{\partial E_{\bf k}{(1)}}
\right]
+O(v^3)\ .
\label{f-d}
\end{equation}
$v$ is the magnitude of ${\bf v}$.
It is clear that to the accuracy of first order in the electric field,
$\widetilde{f}_{\bf k}$ coincides with
$f_{\bf k}^{d}$. Their difference is given by the right hand side of
(\ref{f-d}) which belongs to the terms of high order in the electric field.
It hints that the displaced Fermi function $f_{\bf k}^{d}$
and hence the
balance equation theory of Lei and Ting\cite{lt} is applicable only
when the electric field is so low that the contribution
from the terms on the right hand side of
(\ref{f-d}) are small and can be neglected.
As an example, we calculate the nonlinear impurity-limited resistance of
an electron gas using either the displaced Fermi function
(\ref{flt}) or the new one (\ref{int_sol}). This problem has been studied
by Lei and Ting \cite{lt2} based on the momentum balance equation and
the distribution function (\ref{flt}). We repeat their calculation
in the strong degenerate limit
and the results are plotted as $R_i/R_{i0}$ versus $v/v_F$ curves
in Fig.~3. The curve $a$ ( $b$ ) is calculated by using
(\ref{int_sol}) ( (\ref{flt}) ). $R_{i0}$ is the resistance $R_i$
at $v=0$. $v_F$ is the Fermi velocity. For $N/V=1\times 10^{16}\mbox{cm}^{-3}$
and $m=0.1m_e$, $v_F=7.7\times 10^4 \mbox{m/s}$.
The impurity scattering
is assumed due to charged impurities with bare potential $v(q)\propto q^{-2}$.
The Coulomb interaction between carriers is weakened due to the
large permittivity $\epsilon$ of the lattices  and takes the
form $\displaystyle\frac{\lower.5ex\hbox{$e^2$}}{\epsilon_0\epsilon q^2}$\cite{lt2}.
$\epsilon=10$.
The two curves shown in Fig.~3 coincide approximately only when
$v<0.2v_F$.
But they are quite different when $v > 0.2v_F$.
Choosing $\tau \sim 1 \mbox{ps}$, the electric field $E$ correspond to
$v \sim 0.2v_F$ is estimated to be $90 \mbox{V/cm}$.

It has been shown by Marchetti and Cai \cite{mc} that the momentum
and energy balance
equations of Lei and Ting
can be derived from the conventional Boltzmann equation (if we
do not concern of the concrete expressions for the distribution
function).
The same argument
is applicable to BEPP if the shift of the band energy $E({\bf k})$ in
the electric field \cite{wu2} is taken into account
in the collision term of the Boltzmann
equation. In view of the achievement of the Boltzmann equation in the
condensed matter physics, we believe that the approach based on BEPP
and (\ref{int_sol}) would give better predication for the hot-electron
transport in the higher electric field.
The detail of the numerical results will be reported in a separate paper.

It is pointed out in \cite{wu2} that our theory is applicable only when the
electric field should not be too higher\footnote{)For example, the Wannier
 level is observed only when the electric field is not lower than
 \mbox{$10^5$ V/cm} for $\tau \sim 1$ps \cite{qtss}.}$^)$ so that the Wannier
 levels and the tunneling between the bands \cite{qtss} can be neglected. In such
 lower electric field, the intracollision field effect
 \cite{xlwd} can also be neglected.

\vskip 3cm
{\noindent\Large\it Acknowledgements}

We would like to thank Dr. M. W. Wu for very valuable discussion.

This work is supported by the Grand LWTZ---1298 of Chinese Academy of
Science. We also thank the Center of Superconducting Research and Development
of China for financial support.

\newpage\appendix
\section*{}
In this appendix, we show that
there is an inconsistency in the theory presented in \cite{wu2}.

The statistical average of the total momentum of electron gas
$
\hat{\bf P}=\sum\limits_{{\bf k}\sigma}{\bf p}_{\bf kk}c_{{\bf k}\sigma}^{+}
c_{{\bf k}\sigma}
$  is given by
\begin{equation}
\overline{\bf P}(t)=\sum_{{\bf k}\sigma} {\bf p}_{\bf kk}
\langle\widetilde{c}_{{\bf k}\sigma}^{+}(t)
\widetilde{c}_{{\bf k}\sigma}(t)\rangle\ .
\label{a1}
\end{equation}
Here $\langle\cdots\rangle=\mbox{tr}(\cdots\hat{\rho}_0)$.
By the method of \cite{wu2}, we express (\ref{a1}) in terms of the 
closed time path Green function as follows:
\begin{equation}
\overline{\bf P}(t)=-i \lim_{t^\prime\rightarrow t+0^+}
\sum_{{\bf k}\sigma}{\bf p}_{\bf kk}\mbox{tr}\!\left(\hat{L}
\underline{G}({\bf k}t, {\bf k}t^\prime)\right)\ .
\label{a2}
\end{equation}
Here, $\underline{G}$ is defined by (17) to (20) of \cite{wu2}. In
steady state,
$\underline{G}({\bf k}t,{\bf k}t^\prime)$ depends only on ${\bf k}$
and $t-t^\prime$ and so $\overline{\bf P}(t)$ is independent on time.
Performing Fourier transformation of $\underline{G}$ with respect to
$t-t^\prime$, we obtain
\begin{equation}
\overline{\bf P}=\sum_{{\bf k}\sigma}{\bf p}_{\bf kk}\int
\frac{d\omega}{2\pi i}e^{i\omega 0^+}\mbox{tr}\!\left(\hat{L}\underline{G}
({\bf k}, \omega)\right)\ .
\label{a3}
\end{equation}
Calculating $\underline{G}$ to first order in the electric field \cite{wu2},
we have
\begin{eqnarray}
\overline{\bf P}&=&\sum_{{\bf k}\sigma}{\bf p}_{\bf kk} \int 
\frac{d\omega}{2\pi i} e^{i\omega 0^+}\mbox{tr}\!\left(\hat{L}\underline{G}^0
({\bf k}, \omega)\right)+ \nonumber\\
&+&\sum_{{\bf k}\sigma}{\bf p}_{\bf kk}
\int\frac{d\omega}{2\pi i} \mbox{tr}\!\left(\hat{L}
\lim_{{\bf k}^\prime\rightarrow{\bf k}}\left[\underline{G}^0({\bf k},\omega)
{\hbox{\LARGE\it v}}^E({\bf k}, {\bf k}^\prime)
\underline{G}^0({\bf k}^\prime, \omega)\right]\right)+ O(E^2)\ ,
\label{a4}
\end{eqnarray}
where $\underline{G}^0$ is defined by (24) to (27) of \cite{wu2}. 
One can easily prove the result
\begin{equation}
\overline{\bf P}=Nm{\bf v}_d+\sum_{{\bf k}\sigma}{\bf p}_{\bf kk}
(-e\tau{({\bf k})}{\bf E})\cdot\nabla_{\bf k}f_{\bf k}+O(E^2)\ ,
\label{a5}
\end{equation}
in which $f_{\bf k}$ is given by (12) of \cite{wu2}.
The self-consistent condition requires that the
drift velocity of the electron gas in the initial state, ${\bf v}_d$, should
be equal to that in the final steady state, $\frac{1}{Nm}\overline{\bf P}$
\cite{wu1}. We see
from (\ref{a5}) with disappointment that this self-consistent condition
does not hold. It implies that there are some improper things involved
in the current balance equation theory.

Examining the theory carefully, we find the inconsistency results from 
we choosing the drifting electron gas as the initial state. In order to keep
the consistency of the  theory, we should abandon such initial state and 
adopt instead the electron gas resting in the laboratory coordinate system
as the initial state. In connection with this, the self-consistent condition
to identify the drift velocity of the initial state with that in the
final state is no longer required. Then the first term on the right hand
side of (\ref{a5}) is equal to zero, and (\ref{a5}) is nothing but the
equation defined the drift velocity of the electron gas
in the final steady state, $\frac{1}{Nm}\overline{\bf P}$. The said
inconsistency no longer appears.


\newpage
\begin{center}
\vskip 3pc
\large\bf FIGURES
\end{center}
\begin{description}
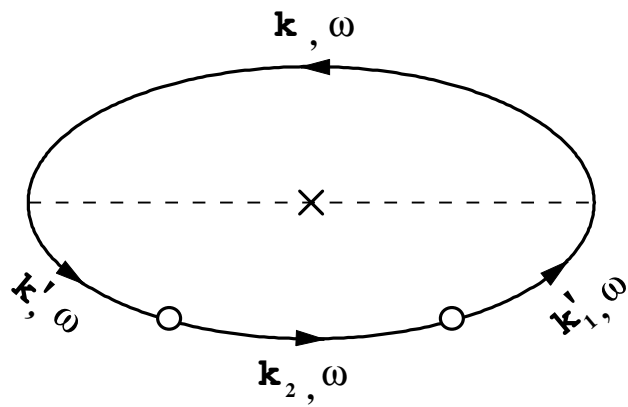
\item[Fig.1] Feynman diagrams for $\underline{G}^\prime$, each with two bare electric
field vertices.
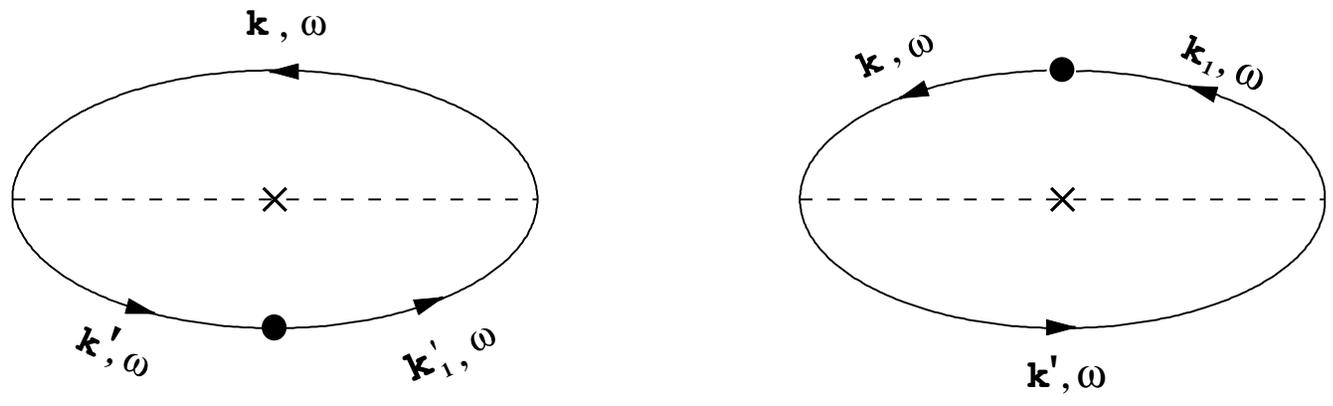
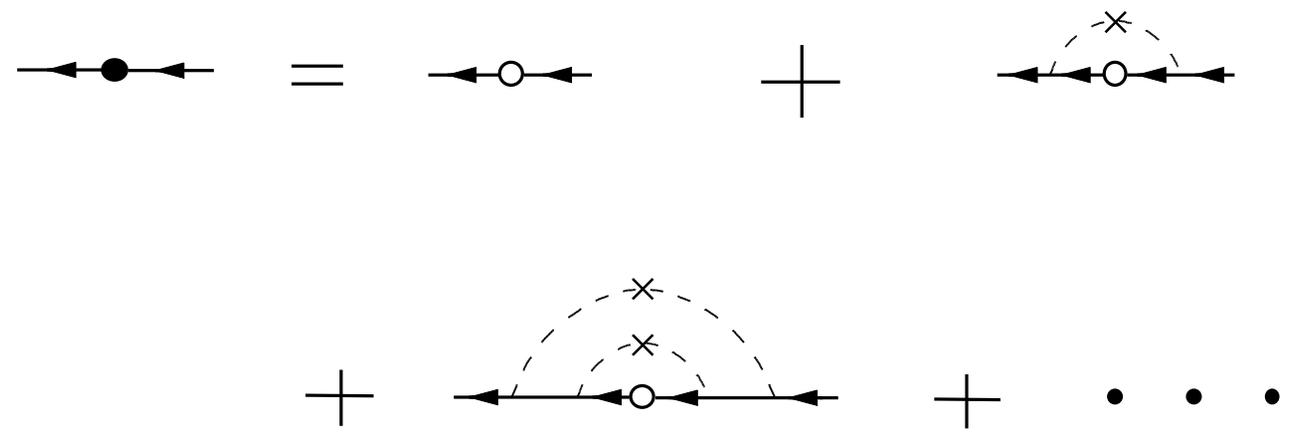
\item[Fig.2] (a) Feynman diagrams for $\underline{G}^\prime$, each with one renormalized
electric field vertex.

\mbox{}\hspace{2mm}(b) Renormalized electric field vertex in the ladder approximation.

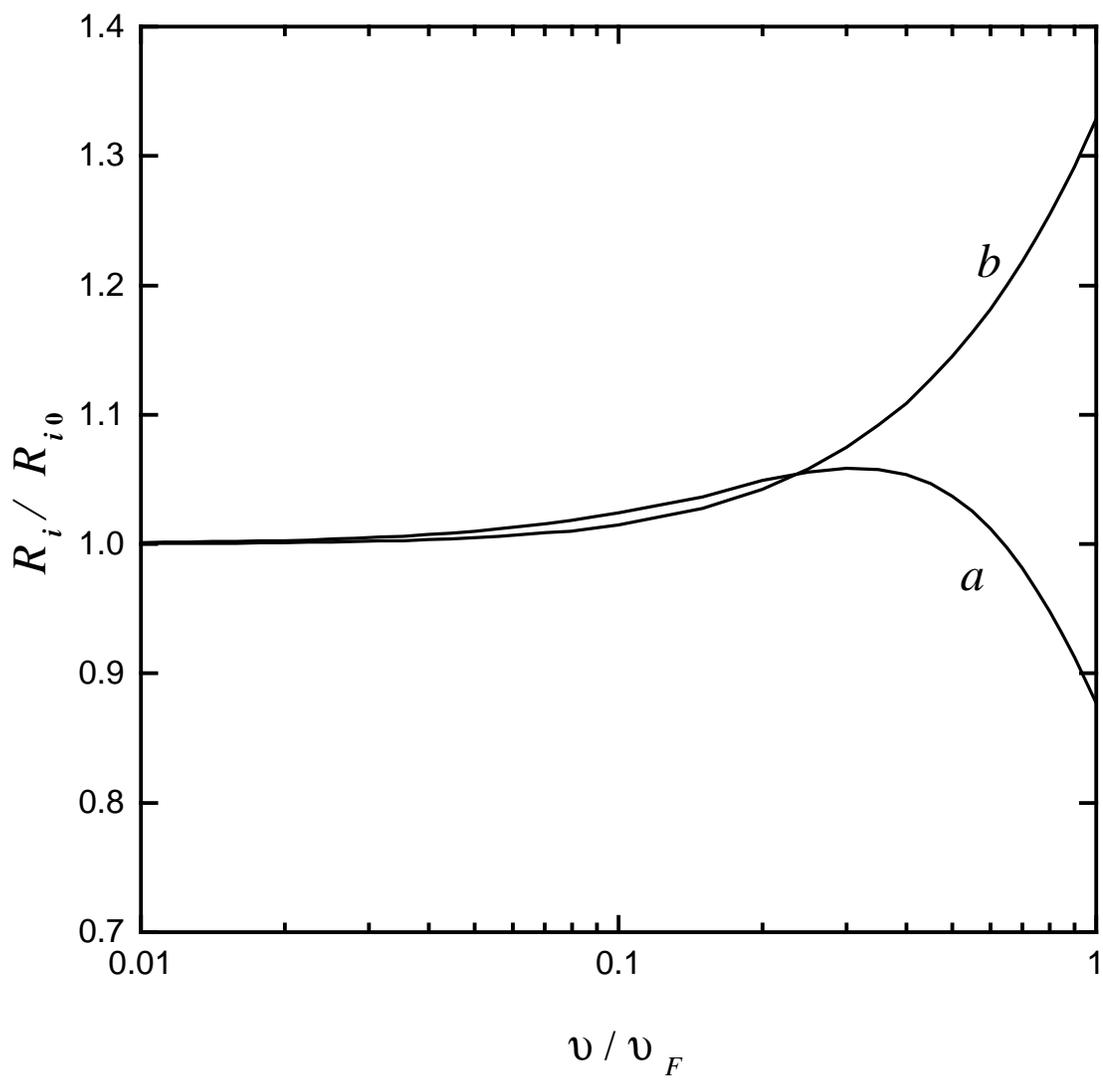
\item[Fig.3] $R_i/R_{i0}$ versus
$v/v_F$ curves.

Curve $a$, using (\ref{int_sol});
curve $b$, using (\ref{flt}).

\end{description}


\begin{references}
\bibitem{wu1}Hang-sheng Wu and M. W. Wu, Phys. Status Solidi (b) {\bf 192}, 129 (1995).
Part 1 of the Series.
\bibitem{lt}X. L. Lei and N. J. M. Horing, Internat. J. Mod. Phys. B {\bf 6}, 805 (1992).
\bibitem{wu2}Hang-sheng Wu, M. W. Wu and Xian-Xiang Huang, Phys. Status Solidi (b) {\bf 198}, 785(1996).
Part 2 of the Series.
\bibitem{cornwell} E.M. Conwell, {\it High Field Transport in Semiconductors},
Academic, New York, 1967.
\bibitem{ab}A. A. Abrikosov, L. P. Gorkov and I. E. Dzaloshinski, 
{\it Methods of Quantum Field Theory in Statistical Physics}, Prentice-Hall,
Englewood Cliffs, N. J., 1963.
\bibitem{mahan}G.D. Mahan, {\it Many-Particle Physics}, Plemum Press, New York, 1981.
\bibitem{lt2} X.L. Lei and C.S. Ting, J. Phys. C: Solid State, {\bf 18},
77(1985).
\bibitem{mc} M. C. Marchetti and W. Cai, Phys. Rev. B {\bf 36}, 8159(1987).
\bibitem{qtss} J. Callaway, {\it Quantum theory of the Solid State}, Section 6,
2nd ed., Academic Press, New York 1991.
\bibitem{xlwd} P. N. Argyres, Phys. Rev. {\bf 117}, 315(1960);
               I. B. Levinson. Sov. Phys. JETP {\bf 30}, 362(1970).
\end{references}
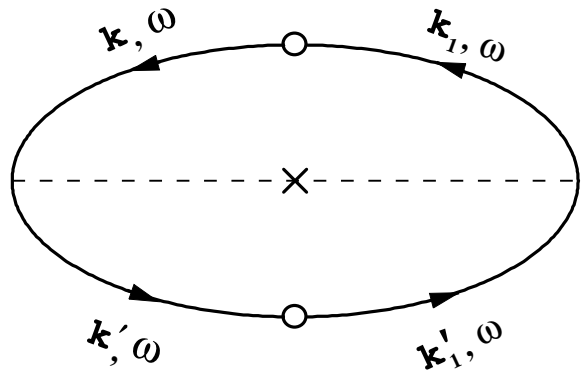
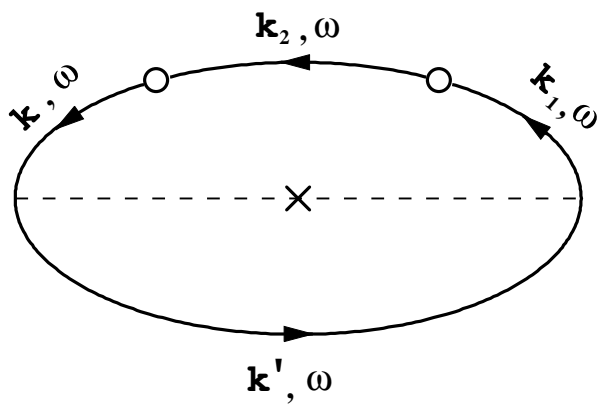
\end{document}